\documentclass[10pt]{article}

\usepackage{a4}
\usepackage{pifont}

\newcommand\atsymbol{@}
\usepackage{amsmath}
\usepackage{amssymb,amsbsy}	

\input{macros2e97}
\newcommand\EXPSPACE{{\rm EXPSPACE}}
\newcommand\NEXPTIME{{\rm NEXPTIME}}
\newcommand\EXPTIME{{\rm EXPTIME}}

\newcommand\PH{{\rm PH}}
\newcommand\CH{{\rm CH}}
\newcommand\PP{{\rm PP}}

\renewcommand\P{{\rm P}}
\newcommand\FP{{\rm FP}}
\newcommand\NP{{\rm NP}}

\newcommand\PSPACE{{\rm PSPACE}}

\newcommand\Sigp[1]{\Sigma^{\rm p}_{#1}}

\renewcommand\L{{\rm L}}
\newcommand\LOGSPACE{{\rm LOGSPACE}}

\newcommand\co{{{\rm co}}}
\newcommand\ex{\exists}

\newcommand\pari{\mathord{\oplus}}
\newcommand\num{\mathord{\rm\#}}
\newcommand\Mod{\mathord{\rm Mod}}

\newcommand\MODPH{\text{\rm MOD-PH}}

\newcommand\NCe{{\rm NC}^1}
\newcommand\ACn{{\rm AC}^0}
\newcommand\ACCn{{\rm ACC}^0}
\newcommand\TCn{{\rm TC}^0}

\newcommand\qNCe{{\rm qNC}^1}
\newcommand\qTCn{{\rm qTC}^0}
\newcommand\qACn{{\rm qAC}^0}

\newcommand\DLOGTIME{{\rm DLOGTIME}}

\newcommand\FDLOGTIME{{\rm FDLOGTIME}}
\newcommand\LOGCFL{{\rm LOGCFL}}
\newcommand\POLYLOGTIME{{\rm POLYLOGTIME}}
\newcommand\FPOLYLOGTIME{{\rm FPOLYLOGTIME}}

\newcommand\PLT{\POLYLOGTIME}

\newcommand\CC{{\mathcalli{C}}}

\newcommand\F{{\mathcalli{F}}}
\newcommand\K{{\mathcalli{K}}}

\newcommand\N{{\mathbb{N}}}	

\newcommand\set[2]{\bigl\{\,#1\bigm| #2\,\bigr\}}
\newcommand\seq{\subseteq}

\newcommand\eqdef{=_{\rm def}}

\newcommand\leafM{{\rm leafstring}^{M}}
\newcommand\leqplt{\leq^{{\it plt}}_m}

\newcommand\C{{\mathcalli{C}}}

\newcommand\quark{\kern.5em}

\newcommand\CSL{{\rm CSL}}
\newcommand\CFL{{\rm CFL}}
\newcommand\REG{{\rm REG}}

\newcommand\mathcalli[1]{\mathcal{#1}}	

\renewcommand\leafM{{\rm leafstring}^{M}}
\newcommand\Leaf{{\rm Leaf}}
\newcommand\BLeaf{{\rm BLeaf}}
\newcommand\LeafM[1]{\Leaf^M(#1)}
\newcommand\Leafp[1]{\Leaf^\P(#1)}
\newcommand\BLeafp[1]{\BLeaf^\P(#1)}
\newcommand\FBTLeafp[1]{{\rm FBTLeaf}^\P(#1)}
\newcommand\BLeafl[1]{{{\rm Leaf}^{\rm L}(#1)}}
\newcommand\BLeaflt[1]{{{\rm Leaf}^{\rm LT}(#1)}}

\newcommand\x[3]{\chi_{#1}[#2\dots #3]}
\newcommand\op[3]{(#1)^{#2}#3}
\newcommand\opp[2]{(#1)^{\rm p}#2}
\newcommand\opl[2]{(#1)^{\rm log}#2}
\newcommand\oppl[2]{(#1)^{\rm plog}#2}
\newcommand\ope[2]{(#1)^{\rm exp}#2}
\newcommand\oppP[1]{(#1)^{\rm p}\P}

\newcommand\FO{{\rm FO}}
\newcommand\SO{{\rm SO}}
\newcommand\calA{{\mathcal{A}}}

\newcommand\CR{{\mathcal{R}}}
\newcommand\Qn{Q_B^{0}}
\newcommand\Qe{Q_B^{1}}

\newcommand\Struct[1]{{\rm Struct}(#1)}
\newcommand\calL{\mathcal{L}}
\def\Neut{\mathcal{N}}

\newcommand\Bs{\{0,1\}^*}
\newcommand\Ss{\Sigma^*}
\newcommand\Gs{\Gamma^*}
\newcommand\Siglt[1]{\Sigma_{#1}^{\rm log}}
\newcommand\Pilt[1]{\Pi_{#1}^{\rm log}}
\newcommand\US{{\rm US}}
\newcommand\toL{\mbox{2-1-$\L$}}

\begin{document}

\title{\Large\bf A Generalized Quantifier Concept in \\
Computational Complexity Theory}

\author{\large\it Heribert Vollmer\\[2ex]\normalsize
Theoretische Informatik\\Universit\"at W\"urzburg\\Am Exerzierplatz 3
\\D-97072 W\"urzburg, Germany\\
vollmer\atsymbol informatik.uni-wuerzburg.de}

\date{}
\maketitle

\begin{abstract}

A notion of generalized quantifier in computational complexity theory
is explored and used to give a unified treatment of leaf language
definability, oracle separations, type 2 operators, and circuits
with monoidal gates. Relations to Lindstr\"om quantifiers are pointed out.


\end{abstract}

\section{Introduction}

In this paper we develop a unified view at some notions that
appeared in computational complexity theory in the past few years.
This will be in the form of operators transforming complexity
classes into complexity classes. Each such operator is given in the form of a
quantifier on strings. This will immediately subsume as special cases
the well known universal, existential, and counting
quantifiers examined in various complexity theoretic settings 
\cite{stme73,wra77,wag86,wag86b,tor91}.
But also a lot of constructions from other subareas of complexity theory
can best be understood in terms of such operators. These include
circuits with arbitrary monoidal gates
\cite{baimst90,baim94}, oracle operators \cite{bawa96,bovowa96},
leaf languages (introduced in \cite{bocrsi92,ver93a} and examined
for different computation models in \cite{helascvowa93,jemcth94,camcthvo96}).
We survey some results from these areas and establish some new connections.

In finite model theory, examinations of the expressive power of various
logics enhanced by Lindstr\"om quantifiers form a very well established
field of active research. Descriptive complexity theory has characterized
a great bulk of complexity classes by such logics. 
We will show that classes defined by our general operator can 
in a uniform way be characterized by model theoretic means using
Lindstr\"om quantifiers.

In the following, we assume some familiarity of the reader with
basic formal language theory (refer to \cite{rosaI97}),
basic complexity classes and resource-bounded reducibilities (refer to
the standard literature, e.g.~\cite{pap94,bocr94,badiga95}; all complexity
classes that appear in this paper without definition
are defined in \cite{joh90}), as well
as with the basics of finite model theory (refer to \cite{vaa94,ebfl95}).

\section{Definition}
\label{defin}

Given a language $A$ over some alphabet $\Sigma$, we denote the
characteristic function of $A$ by $\chi_A$, i.e. for all $x\in\Ss$,
$\chi_A(x)=1$ if $x\in A$, and $\chi_A(x)=0$ otherwise.

We will always assume some order on the alphabets we use; therefore
it makes sense to talk about the lexicographic order $\prec$ of $\Ss$,
and for $x,y\in\Ss$, $x\preceq y$, we define the characteristic string
of $A$ from $x$ to $y$ as 
$\x{A}{x}{y}\eqdef\chi_A(x)\chi_A(x+1)\cdots\chi_A(y)$. 
Here, $x+1$ denotes the successor of $x$. In fact, we will presuppose
an underlying bijection between $\Ss$ and the set $\N$ of natural numbers,
and we use the notation $\x{A}{i}{j}$ for $i,j\in\N$. 

Let $\langle\cdot,\cdot\rangle$ denote a standard pairing function.
For a set $A\seq\Ss$ and a string $x\in\Ss$, define
$A_x\eqdef\set{y}{\langle x,y\rangle\in A}$.

Looking now at the well-known characterization of the polynomial hierarchy by 
polynomially length-bounded universal and existential quantifiers 
$\ex^p$, $\forall^p$ \cite{wra77},
the following is clear:

\begin{itemize}
\item A language $L$ is in $\NP$ if and only if there is a language $A\in\P$
and a function $f$ computable in polynomial time such that for all $x$,
$$x\in L \iff \x{A_x}{0}{f(x)} \in (0+1)^*1(0+1)^*
\ \text{(i.e., contains a ``$1$'')}.$$
\item A language $L$ is in $\co\NP$ if and only if there is a language $A\in\P$
and a polynomial $p$ such that for all $x$,
$$x\in L \iff \x{A_x}{0}{f(x)} \in 1^*
\quad\text{(i.e., consists out of ``$1$''s only)}.$$
\end{itemize}

An analogous result holds for the class $\PP$, which was
characterized in \cite{wag86b} in terms of the so called 
polynomially length-bounded counting quantifier ${\rm C}^p$:

\begin{itemize}
\item A language $L$ is in $\PP$ if and only if there is a language $A\in\P$
and a function $f$ computable in polynomial time such that for all $x$,
$$x\in L \iff \x{A_x}{0}{f(x)}
\text{ contains more ``$1$''s than ``$0$''s}.$$
\end{itemize}

The class $\US$ (for unique solution) is defined by polynomial time 
nondeterministic Turing machines $M$ which accept an input $x$ if and only
if there is exactly one accepting path in the computation tree of $M$ on $x$.

\begin{itemize}
\item A language $L$ is in $\US$ if and only if there is a language $A\in\P$
and a function $f$ computable in polynomial time such that for all $x$,
$$x\in L \iff \x{A_x}{0}{f(x)} \in 0^*10^*.$$
\end{itemize}

Thus we see that here the semantics of quantifiers is defined by giving
languages over the binary alphabet 
($E\eqdef(0+1)^*1(0+1)^*$ for $\ex^p$, $U\eqdef1^*$ for $\forall^p$, and
${\rm maj}\eqdef\set{w\in\Bs}{w \text{ contains more ``$1$''s than ``$0$''s}}$
for ${\rm C}^p$). The following generalization now is immediate:

Let $B\in\Bs$, let $\K$ be a class of sets, and $\F$ be a class
of functions from $\Sigma^*\rightarrow\N$. Define the class
$\op{B}{\F}{\K}$ to consists of all sets $L$ for which there exist
some $A\in\K$ and some function $f\in\F$ such that 
for all $x\in\Ss$, $x\in L \iff \x{A_x}{0}{f(x)}\in B$.

We use the following shorthands: 
Write $\opp{A}{\K}$ ($\opl{A}{\K}$, $\oppl{A}{\K}$, resp.), 
if $\F$ is the class of all functions from $\Sigma^*\rightarrow\N$ 
computable in polynomial time (logarithmic time, polylogarithmic time, resp.) 
on deterministic Turing machines 
(i.e., $\F=\FP$, $\F=\FDLOGTIME$, $\F=\FPOLYLOGTIME$, resp.). 
For sub-linear time bounds we use Turing machines with index tape and random
access to their input, working in the unrestricted mode (for background,
refer to \cite{cach95,revo97}). Observe that a function $f\in\FP$ is 
polynomially length-bounded, $f\in\FDLOGTIME$ is length-bounded by some
function $c\cdot\log n$, and $f\in\FPOLYLOGTIME$ is polylogarithmically 
length-bounded.
If $\cal L$ is a class of languages, then 
$\op{{\cal L}}{\F}{\K}\eqdef\bigcup_{B\in{\cal L}}\op{B}{\F}{\K}$.

If we take the above three languages $E$, $U$, and maj, and look at different
function classes $\F$, we get the existential, universal, and
counting quantifier for various length-bounds.

The above definition appeared in \cite{vol96b} and (for the special
case $\F=\FP$) in \cite{bosi97}.

\section{Polynomial Time Leaf Languages}
\label{polyn}

The most examined special case of our general operator is probably
the polynomial time case, i.e.~the base class $\K$ is the class $\P$
(and $\F=\FP$).
In this case there is a very intuitive way of visualizing the operator
via so called {\em leaf languages}.

\subsection{Definition}
\label{defleaf}

In the leaf language approach to the characterization of complexity
classes, the acceptance of a word input to a nondeterministic machine
depends only on the values printed at the leaves of the computation
tree. To be more precise, let $M$ be a nondeterministic Turing
machine, halting on every path printing a symbol from an alphabet
$\Sigma$, with some order on the nondeterministic choices.  Then,
$\leafM(x)$ is the concatenation of the symbols printed at the leaves
of the computation tree of $M$ on input $x$ (according to the order of
$M$'s paths given by the order of $M$'s choices).  
Given now a language $B\seq\Bs$, we define 
$\LeafM{B} = \set{x}{\leafM(x)\in B}$.

Call a computation tree
of a machine $M$ {\em balanced}, if all of its computation paths have
the same length, and moreover, if we identify every path with the
string over $\{0,1\}$ describing the sequence of nondeterministic
choices on this path, then there is some string $z$ such that all
paths $y$ with $|y|=|z|$ and $y\preceq z$ (in lexicographic ordering)
exist, but no path $y$ with $y\succ z$ exists.  

A leaf language $B\seq\Ss$ now defines
the class $\BLeafp{B}$ of all languages $L$ for
which there exists a nondeterministic polynomial time machine $M$
whose computation tree is always balanced, such that 
$L=\LeafM{B}$.
Let $\C$ be a class of languages. The class
$\BLeafp{\C}$ consists of the union over all $B \in \C$ of the
classes $\BLeafp{B}$.

This computation model was introduced by Bovet, Crescenzi, and
Silvestri, and independently Vereshchagin 
\cite{bocrsi92,ver93a} and later examined by Hertrampf,
Lautemann, Schwen\-tick, Vollmer, and Wagner \cite{helascvowa93}, and Jenner,
McKenzie, and Th\'erien \cite{jemcth94}, among others. See also
the textbook \cite[pp.~504f]{pap94}. 

Jenner, McKenzie,
and Th\'erien also considered the case where the computation trees
are not required to be balanced. For that case, let $B$ be any
language. Then, the class $\Leafp{B}$ consists of those languages $L$
for which there exists a nondeterministic polynomial time machine $M$
without further restriction, such that $L=\LeafM{B}$.
Let $\C$ be a class of languages. The class $\Leafp{\C}$ consists of
the union over all $B \in \C$ of the classes $\Leafp{B}$.
(Strictly speaking, the
definition of {\em balanced\/} given in \cite{jemcth94} is different
from ours and, at first sight, slightly more general. However, it is
easy to see that both definitions are equivalent.) 

The reader now might wonder about the seemingly unnatural condition that
the nondeterministic choices of $M$ are ordered. In fact, most complexity
classes of current focus can be defined without this assumption---in this
case the leaf language $B$ has the special property that we can permute
the letters in a given word without affecting membership in $B$. (Cf.~our
results on {\em cardinal languages\/} in Sect.~\ref{oracle} below.)
However, strange classes where the order of the paths is important for
their definition are conceivable, and the results presented below,
especially the oracle separation criterion (Theorem~\ref{septhm}),
also hold for these pathologic cases.

The following connection to our generalized quantifier now is not
too hard to see.

\begin{thm}\label{releasy}
Let $B\seq\Bs$. Then $\oppP{B} = \BLeafp{B}$.
\end{thm}

\begin{outline}
($\seq$) Let $L\in\oppP{B}$, $x\in L\iff \x{A_x}{0}{f(x)}\in B$.
The nondeterministic machine, given $x$, branches on all possible
second inputs $y$ in the range $0,\dots,f(x)$, and outputs $\chi_A(x,y)$.

($\supseteq$) Let $L\in\BLeafp{B}$ via the nondeterministic machine $M$.
Computation paths of a nondeterministic machines can be followed in 
polynomial time if the nondeterministic choices are known. 
Defining $A$ to consist of all pairs $(x,p)$ such that
$p$ is a sequence of nondeterministic choices leading to a 
path of $M$ that outputs ``$1$'' and $f(x)$ to be the number of
paths of $M$ on input $x$, we have $x\in L\iff \x{A_x}{0}{f(x)}\in B$.
\end{outline}

The definition of leaf languages allows for languages $B$ not necessarily
over the binary alphabet. If we want to come up with a connection to our
generalized quantifier also for such $B$, we face a
problem. In the definition in Sect.~\ref{defin} the binary alphabet seems essential.
Fortunately, for every $B$ there is usually a $B'\seq\Bs$ such that
$\BLeafp{B}=\BLeafp{B'}$, where $B$ and $B'$ are of the same complexity.
In most cases, $B'$ can simply be obtained from $B$ by block encoding
(then $B$ and $B'$ are $\FO$-equivalent).
We come back to this point in the next subsection.

\subsection{The Complexity of a Leaf Language}

In \cite{helascvowa93} the question how complex a leaf language must be
in order to characterize some given complexity class $\K$ was addressed.
Let us start by considering some examples.

At great number of classes can be defined by regular leaf languages.
This is obvious for 
$\NP$, $\co\NP$ and $\US$ as we saw in the previous section,
for $\Mod_k\P$ (all words with a number of ``$1$''s divisible by $k$),
but also true for higher levels of the polynomial hierarchy (see below) and
the boolean hierarchy over $\NP$ (e.g.~the class $\NP\wedge\co\NP$ can be
defined via the set of all words such that the string ``$010$'' appears at
least once, but the string ``$0110$'' does not appear).

For other complexity classes, context-free languages come immediately to mind.
$\PP$ can obviously be defined by the language maj from the previous section.
Recalling the characterization of $\PSPACE$
via polynomial time alternating Turing machines, it is clear that the
set of all (suitably encoded) boolean expressions involving the constants
``true'' and ``false'' and the connectives AND and OR that evaluate
to ``true'', is an appropriate leaf language. 

The question however arises if we can do better here. It was shown
in \cite{helascvowa93}, that in the case of $\PSPACE$ there is a regular
leaf language.

Let $S_5$ denote the word problem for the group of permutations on five
elements (suitably encoded over the binary alphabet), i.e.~$S_5$ consists
of sequences of permutations which multiply out to the empty permutation.

\begin{thm}\label{pspace}
$\oppP{S_5}=\PSPACE$.
\end{thm}

\begin{outline}
For the inclusion from left to right, just observe that a $\PSPACE$ machine
can traverse the whole computation tree of a given nondeterministic machine
to evaluate the product over $S_5$. This simulation then stops accepting if
and only if the result is the identity permutation.

For the other direction, we are given a language $L\in\PSPACE$.
Then there is a polynomial time alternating Turing machine accepting $L$.
Thus, for every input $w$, 
machine $M$ defines a polynomial depth computation tree $T(w)$
where the leafs carry values $0$ or $1$ and in the inner nodes the functions
AND and OR are evaluated. $w\in L$ iff the root of this tree evaluates to $1$.
As a first step we transform this tree into a tree $T'(w)$ where in all the
inner nodes the function NOR is evaluated. This can easily be achieved since
the NOR function constitutes a complete basis for the boolean functions.

As a second step we now ``simulate'' NOR in $S_5$. This simulation is 
essentially due to David Barrington \cite{bar89}. 
Let $b, c, d, e, f$ be the following permutations from the group $S_5$:
$$b=(23)(45), c=(12435), d=(243), e=(345), f=(152)$$
Further let $a_0$ be the empty permutation, denoted by $a_0=()$, and
let $a_1 = (12345)$. Now consider the following product in $S_5$ including the 
variables $x$ and $y$:
$$w(x,y)=a_0bx^4cy^4dxeyf$$
Simple calculations show that 
$w(a_0,a_0)=a_1$ and $w(a_0,a_1)=w(a_1,a_0)=w(a_1,a_1)\allowbreak=a_0$.
Thus coding the value {\em true\/} by $a_1$, and {\em false\/} by $a_0$,
we can view $w$ as the {\rm NOR}-operation applied to $x$ and $y$.

Now replace every appearance of a ``NOR''-node in $T'(x)$ with sons $x$ and 
$y$ by a binary subtree of height 4 whose 16 leaves are
$$a_0~~b~~x~~x~~x~~x~~c~~y~~y~~y~~y~~d~~x~~e~~y~~f$$

Thus we accept the input $w$ if and only if the leaf string evaluates to 
$a_1$. Taking $B$ to be
the regular language 
$$B\eqdef\set{x}{x \text{ is a string of elements from $S_5$
which evaluates to $a_1$}},$$
we then get $\Leafp{B} = \PSPACE$. 
It is easy to go from $B$ to the word problem $S_5$ (by just adding one
more factor $a_1^{-1}$), and since we have an identity element which we
can insert arbitrarily in the leaf string to fill gaps in the computation tree
in order to make it balanced, we get 
$\PSPACE = \Leafp{B} = \BLeafp{S_5} = \oppP{S_5}$. 
\end{outline}

The question now of course is what is so special about the language
$S_5$. What can be said more generally? Using deep algebraic
properties of regular languages exhibited in \cite{the81,bath88} 
(see also the textbook \cite{str94}) one can
show the following.

Let $\PH$ denote the union of all classes of the polynomial hierarchy
\cite{stme73},
i.e.~$\PH=\NP\cup\NP^\NP\cup\NP^{\NP^\NP}\cup\cdots$.
Let $\MODPH$ denote the oracle hierarchy constructed similarly, but now 
allowing
as building blocks not only $\NP$ but also all classes $\Mod_k\P$ for
arbitrary $k\in\N$.

\begin{thm}[\cite{helascvowa93}]\label{algebraic}
\begin{enumerate}
\item 
Let $A$ be a regular language whose syntactic mo\-no\-id is non-solvable.
Then $\oppP{A}=\PSPACE$.
\item 
Let {\rm SOLVABLE} denote the class of all regular languages whose syntactic 
monoid is solvable. Then $\oppP{{\rm SOLVABLE}}=\MODPH$.
\item 
Let {\rm APERIODIC} denote the class of all regular languages whose syntactic
monoid is aperiodic. Then $\oppP{{\rm APERIODIC}}=\PH$.
\end{enumerate}
\end{thm}

Regular leaf languages for individual levels of the polynomial hierarchy
can also be given. For example $\Sigp{2}$ can be defined over 
$\Sigma=\{a,b,c\}$ by $\Sigma^*ca^+c\Sigma^*$, intuitively: ``there is
a block consisting out of `$a$'s only''. This is an $\exists\forall$
predicate directly reflecting the nature of $\Sigp{2}$-computations.
If we now chose a simple block encoding this might lead us out of the
aperiodic languages. However, we may proceed as follows: Define
$A_2=(0+1)^*11(010)^+11(0+1)^*$. It is clear that this leaf language defines
a subclass of $\Sigp{2}$---just check that there are two substring $11$ such
that in between we have a sequence of occurrences of the 3-letter string
$010$; this is an $\exists\forall$ condition. On the other hand,
suppose we are
given a $\Sigp{2}$ machine $M$, i.e.~an alternating machine with computation
trees consisting of one level of $\exists$ nodes followed by a second level of
$\forall$ nodes; i.e. the initial configuration is the root of an
existential tree where in the leaves we append universal subtrees.
We transform this into a tree where we use the substring
``$11$'' in the leafstring
as separator between different $\forall$ subtrees, and within each
such subtree we simulate an accepting path by the 3 leaf symbols ``$010$'' 
and a rejecting path by the symbol ``$0$''. 
Then $M$ produces a tree with at least one
universal subtree consisting out of only accepting paths 
iff the leaf word of this simulation is in $A_2$.
$\Sigp{3}$ can similarly be defined via
$A_3=(0+1)^*111\overline{A_2}111(0+1)^*$. This generalizes to 
higher levels of the polynomial hierarchy. 
With some care one can show that $A_2$ and $A_3$ are
in levels ${\cal B}_2$ and ${\cal B}_3$, resp.,
of the Brzozowski-hierarchy of regular languages.
This hierarchy of star-free regular languages measures the nesting depth
of the dot (i.e.~concatenation) operation. For a formal definition see
\cite{eil76}.
More generally the following holds:

\begin{thm}[\cite{helascvowa93}]
$\oppP{{\cal B}_k}$ is the boolean closure of the class ${\Sigp{k}}$.
\end{thm}

Let us now come back to the question if $\PP$ (for which we gave
a context-free leaf language above) can also be done by a regular language.

\begin{coro}
$\PP$ is not definable via a regular leaf language unless either
$\PP=\PSPACE$ or $\PP\seq\MODPH$.
\end{coro}

\begin{proof}
If there is a regular leaf language $L$ for $\PSPACE$, then there
are two cases to consider: either $L$ is non-solvable (in this case
$\PP=\PSPACE$) or $L$ is solvable (then $\PP\seq\MODPH$).
\end{proof}

In \cite{helascvowa93} leaf languages defined by restricting
resource bounds as time and space were examined. It was shown that
the complexity class obtained in this way is defined via the same
resource, but the bound is one exponential level higher, for
example $\oppP{\P}=\EXPTIME$, $\oppP{\NP}=\NEXPTIME$, 
$\oppP{\LOGSPACE}=\PSPACE$, $\oppP{\PSPACE}=\EXPSPACE$, and so on. 
Denoting the levels
of the alternating log-time hierarchy \cite{sip83b} by $\Siglt{k}$ ($k\in\N$),
we get the following special case:

\begin{thm}\label{pollogsigma}
$\oppP{\Siglt{k}} = \Sigp{k}$.
\end{thm}

\subsection{Some Complexity Theoretic Applications}

\subsubsection{Normal Forms}
\label{normal}

The characterization of $\PSPACE$ (Theorem~\ref{pspace}) was somewhat
surprising, since it points out a very restricted normal form for
$\PSPACE$ computations. Cai and Furst defined a class $\K$ to be
{\em $\K'$-serializable}, if every $\K$ computation can be organized
into a number of local computations $c_1,\dots,c_r$ 
(which in turn are restricted to be $\K'$ computations),
each passing only a constant number $k$ 
of bits as the result of its computation to the
next local computation. 
The sequence $c_1,\dots,c_r$ is uniform in the sense that there is one
$\K'$ program that gets as
input only the original input, a number $i$, and a string of $k$ bits, 
and computes the $k$-bit-result of $c_i$'s computation.
Please refer to \cite{cafu91} for a formal definition. Machines as
just described are also called bottleneck machines. The {\em bottleneck\/}
refers to the restricted way of passing information onwards.

\begin{coro}[\cite{helascvowa93}]\label{pspaceserial}
$\PSPACE$ is $\ACn$-serializable.
\end{coro}

\begin{outline}
Let $L\in\BLeafp{S_5}$ via machine $M$.
The information passed from one computation to the next will be an
encoding of an element of the group $S_5$.
Each local computation uses its number to recover from it a path of the
nondeterministic Turing machine. (If the number does not encode a
correct computation path, then we simply pass the information
we get from our left neighbor onwards to the right.) The leaf
symbol on this path is then multiplied to the permutation we got
from the left, and the result is passed on to the right.
This can be done in $\ACn$ since computation paths of polynomial time
Turing machines can be
checked in $\ACn$. (A computation path consists not only out of $M$'s
nondeterministic choices, but is a complete sequence of configurations
of $M$.)
\end{outline}

The power of bottleneck machines was examined in detail in \cite{her97}.
He gave a connection between these machines and leaf languages defined
via transformation monoids. The power
of bottleneck machines as a function of the number of bits passed
from one local computation to the next was determined.

\subsubsection{Oracle Separations}
\label{oracle}

The original motivation for the introduction of leaf languages in
\cite{bocrsi92,ver93a} was the wish to have a uniform oracle separation
theorem. Usually when relativized complexity classes are separated, this
is achieved by constructing a suitable oracle by {\em diagonalization,}
usually a stage construction. Bovet, Crescenzi, Silvestri, and 
Vereshchagin wanted to identify the common part of all these constructions
in a unifying theorem, such that for future separations, one could concentrate
more on the combinatorial questions which are often difficult enough.
They showed that to separate two classes defined by leaf languages,
it is sufficient to establish a certain non-reducibility between the
defining languages. Let $A,B\seq\Bs$. Say that $A$ is polylogarithmic time
bit-reducible to $B$, in symbols: $A\leqplt B$, 
if there are two functions $f,g$ computable in 
polylogarithmic time such that for all $x$, 
$x\in A\iff f(x,0)f(x,1)\cdots f(x,g(x))\in B$.

\begin{thm}[\cite{bocrsi92,ver93a}]\label{septhm}
Let $A,B\seq\Bs$. Then $A\leqplt B$ if and only if for all oracles $Y$,
the inclusion $\oppP{A}^Y\seq\oppP{B}^Y$ holds.
\end{thm}

Observe that $A\leqplt B$ is just another formulation for the containment
of $A$ in $\oppl{B}{\PLT}$, which in turn is equivalent to 
the inclusion of the class $\oppl{A}{\PLT}$ in $\oppl{B}{\PLT}$.

\begin{coro}\label{plogs}
Let $A,B\seq\Bs$. Then we have:
$$\oppl{A}{\PLT}\seq\oppl{B}{\PLT}$$
if and only if for all oracles $Y$,
the inclusion 
$$\oppP{A}^Y\seq\oppP{B}^Y$$
holds.
\end{coro}

In \cite{bosi97}, Theorem~\ref{septhm} was strengthened as follows: 
It was shown
that $\opp{A}{\K}\seq\opp{B}{\K}$ for all nontrivial classes $\K$ if and only
if $A$ is reducible to $B$ by monotone polyloga\-rith\-mic-time uniform
projection reducibility. Refer to their paper for details.

Observe that a polylogarithmic time bit-reduction cannot (simply because
of its time bound) read all of its input. This often allows one to prove
$A\not\leqplt B$ by an adversary arguments. We give a very simple
example.

\begin{example}
Let $E=(0+1)^*1(0+1)^*$, $U=1^*$ as in Sect.~\ref{defin}. 
Then $\oppP{E}=\NP$ and $\oppP{U}=\co\NP$.
Suppose $U\leqplt E$. The input $x=1^n$ must be mapped by this reduction to
a word with at least one ``$1$''. The computation leading to this ``$1$''
however cannot read all of $x$. If we now define $x'$ by complementing in $x$ 
a bit which is not queried, then again $x'$ will be mapped to a string in $E$,
which is a contradiction. Thus $U\not\leqplt E$, and hence there is an 
oracle separating $\co\NP$ from $\NP$.
\end{example}

Vereshchagin in \cite{ver93a} used Theorem~\ref{septhm} to establish
all relativizable inclusions between a number of prominent classes within
$\PSPACE$. His list contains besides the classes of the polynomial time
hierarchy also UP, FewP, RP, BPP, AM, MA, PP, IP, and others.

A very satisfactory application of Theorem~\ref{septhm} was
possible in the following special case. Say that $L\seq\Ss$ is a
{\em cardinal language}, if membership in $L$ only depends on the
frequency with which the elements of $\Sigma$ appear in words.
This means that if $\Sigma=\{a_1,\dots,a_k\}$ we can associate
$L$ with a set $N(L)\seq\N^k$, in such a way that $w\in L$ iff
there is a $(v_1,\dots,v_k)\in N(L)$ where $a_i$ occurs in $w$
exactly $v_i$ times ($1\leq i\leq k$).
($N(L)$ is the image of $L$ under the Parikh mapping: 
$N(L)=\Psi_\Sigma(L)$.)
Say that $L$ is of bounded significance if there is a number $m\in\N$
such that for all $(v_1,\dots,v_k)$ we have
$$(v_1,\dots,v_k)\in N(L) \iff (\min(v_1,m),\dots,\min(v_k,m))\in N(L).$$
Using Ramsey theory, Hertrampf in \cite{her95b} proved the following:

\begin{thm}[\cite{her95b}]
There is an algorithm that, given two cardinal languages $A,B$ of 
bounded significance, decides if $\oppP{A}^Y\seq\oppP{B}^Y$ for all oracles
$Y$.
\end{thm}

Pushing his ideas just a bit further, the following was proved:
We say that $p\colon\N^k\rightarrow\N$ is a positive linear combination
of multinomial coefficients if 
$p(\vec{v})=\sum_{u\leq z}\alpha_u\binom{v}{u}$ 
for some
$z\in\N^k$, $\alpha_u\in\N$ (for $u\leq z$, the order taken component-wise).

\begin{thm}[\cite{crhevowa97}]\label{multinom}
Let $A,B$ be cardinal languages of bounded significance over a $k$ element
alphabet. Then $A\leqplt B$ if and only if there are functions
$p_1,\dots,p_k\colon\N^k\rightarrow\N$ which are positive linear combinations
of multinomial coefficients, such that for all $\vec{v}=(v_1,\dots,v_k)$,
$\vec{v}\in N(A)$ if and only if 
$(p_1(\vec{v}),\dots,p_k(\vec{v}))\in N(B)$.
\end{thm}

In other words, if such $k$ functions do not exist, then there is an oracle
separating $\oppP{A}$ from $\oppP{B}$.
Thus we see that the oracle separation criterion Theorem~\ref{septhm}
leads to a very strong statement in the context of cardinal languages.
This result was used in \cite{crhevowa97} to establish a complete list
of all relativizable inclusions between classes of the boolean hierarchy
over $\NP$ and other classes defined by cardinal languages of bounded
significance.

Valiant's counting class $\num\P$ is of course strongly related to
the notion of cardinal languages. In the case of $\num\P$ we just deal
with the binary alphabet, and we count the number of ``$1$''s in a
leaf string. Closure properties of $\num\P$, that is operations that don't
lead us out of the class, play an important role to establish inclusions
between complexity classes; e.g.~Toda's result $\PH\seq\P^\PP$ \cite{tod91} 
and Beigel, Reingold, and Spielman's proof that $\PP$ is closed under 
intersection \cite{beresp91} both heavily build on the fact that $\num\P$
is closed under certain sums, products, and choose operations.

Similar to Theorem~\ref{multinom} one can obtain the following:

\begin{thm}[\cite{hevowa95}]
A function $f:\N^k\rightarrow\N$ is a relativizable closure property of 
$\num\P$ (i.e., relative to all oracles, if $h_1,\dots,h_k\in\num\P$ then also
$f(h_1,\dots,h_k)\in\num\P$), if and only if $f$ is a positive linear 
combinations of multinomial coefficients.
\end{thm}

\subsubsection{Circuit Lower Bounds}

Circuit classes as leaf languages have been considered in 
\cite{camcthvo96,vol96}. 
For background on circuit complexity, we refer the reader to
\cite{str94}.
It is immediate from Theorem~\ref{pspace} that 
$\oppP{\NCe}=\PSPACE$. Additionally one can prove e.g.~that
$\oppP{\ACn}=\PH$, and that $\oppP{\TCn}$ is the counting hierarchy
$\CH$, defined in \cite{wag86,wag86b}
as $\PP\cup\PP^\PP\cup\PP^{\PP^\PP}\cup\cdots$.
Finer results are given in \cite{vol97}.

Building on leaf language characterizations, the circuit class 
$\TCn$ (where we require logtime uniformity)
was separated from the counting hierarchy in \cite{camcthvo96}. 
This was improved by Allender \cite{all96} to the following separation.

\begin{thm}
$\TCn\neq\PP$.
\end{thm}

\begin{outline}
We sketch the proof of the weaker result from \cite{camcthvo96}.
Suppose that $\TCn = \CH$. Then we have
$\TCn = \CH = \BLeafp{\TCn} = \BLeafp{\CH} \supseteq \EXPTIME$, thus
$\P\supseteq\EXPTIME$, which is a contradiction.
Allender now observed that this can be extended to show that any language
complete for $\PP$ under $\TCn$ reductions cannot be in $\TCn$.
\end{outline}

In the non-uniform case no similar lower bound for $\TCn$ is known.
If we relax the uniformity condition just a little bit, we know that
$$\oppP{\mbox{\rm logspace-uniform }\ACn}=\PSPACE$$
(thus also
$\BLeafp{\mbox{\rm logspace-uniform }\TCn}=\PSPACE$). This shows
that logtime-uniformity is critical in the above proof.

In Corollary~\ref{plogs} it became clear that
the oracle separability of two polynomial time classes is equivalent to
the absolute separability of two lower classes with the same acceptance
paradigm. A similar relation is known between polynomial time and constant
depth circuit classes. E.g.~building on previous work by Furst, Saxe, and
Sipser \cite{fusasi84}, Yao in his famous paper used a lower bound
for the parity function to construct an oracle separating $\PSPACE$ from
the polynomial hierarchy \cite{yao85}. This connection has been exploited
a number of times since then. 

The formal connection between Theorem~\ref{septhm} and the
Furst, Saxe, Sipser approach to oracle construction has been given in
\cite{vol97}.\nocite{vol96}
The main observation that has to be made is that $\leqplt$-reductions
can be performed by (uniform) $\qACn$ circuits.
$\qACn$ stands for quasipolynomial $\ACn$ \cite{bar92}, i.e.~unbounded
fan-in circuits of constant depth and size $2^{\log^{O(1)}n}$.
(Similarly we will also use $\qTCn$ for quasipolynomial size $\TCn$ circuits,
and $\qNCe$ for quasipolynomial size $\NCe$ circuits.)

\begin{thm}\label{circora}
Let $A,B\seq\Bs$. Then we have:
$A\not\in\oppl{B}{\qACn}$ 
if and only if $\oppl{A}{\qACn}\not\seq\oppl{B}{\qACn}$
if and only if there is an oracle $Y$ such that
$\opp{A}{\PH}^Y\not\seq\opp{B}{\PH}$.
\end{thm}

This theorem can be used to attack the
``nagging question'' \cite{for97}
how to separate superclasses of $\P^\PP$ from $\PSPACE$.
Some special cases are the following.

\begin{coro}
$S_5\not\in\qTCn$ if and only if 
$\qTCn\neq\qNCe$
if and only if there is an oracle separating the
counting hierarchy from $\PSPACE$.
\end{coro}

\begin{outline}
Under the assumption $S_5\in\qTCn$, the following inclusion chain holds
relativizably:
$$\PSPACE = \BLeafp{S_5}\seq\BLeafp{\qTCn}=\CH.$$ 
This proves the direction from right to left.
For the other direction, if relative to all oracles
$\PSPACE\seq\CH$ then $S_5$ polylogarithmic time bit-reduces to $\qTCn$,
but this class is even closed under $\qACn$ reductions.
\end{outline}

Define 
${\rm par}\eqdef\set{w\in\Bs}{\text{the number of ``$1$''s in $w$ is odd}}$,
and let maj be as in Sect.~\ref{defin}. 
   
\begin{coro}
$S_5\not\in\oppl{{\rm maj}}{\oppl{{\rm par}}{\qACn}}$
if and only if there is an oracle separating 
$\PP^{\pari\P}$ from $\PSPACE$.
\end{coro}

\begin{outline}
If $\PSPACE\seq\PP^{\pari\P}$ then $S_5$ polylogarithmic time bit-reduces
to a language in the class $\oppl{{\rm maj}}{\oppl{{\rm par}}{\qACn}}$, and therefore $S_5$ is even
in this class (it is closed under $\leqplt$). 

On the other hand, if
$S_5\in\oppl{{\rm maj}}{\oppl{{\rm par}}{\qACn}}$, then 
$\PSPACE = \BLeafp{S_5}\allowbreak
\seq \BLeafp{\oppl{{\rm maj}}{\oppl{{\rm par}}{\qACn}}}
= \PP^{\pari\P^\PH} = \PP^{\pari\P}$.
\end{outline}

A refinement of Theorem~\ref{circora} and further investigations along
these lines can be found in \cite{vol97}.

\subsection{Definability vs.~Tree Shapes}

Our quantifier from Sect.~\ref{defin} coincides as we saw in
the polynomial time context
with leaf languages for balanced computation trees. The unbalanced
case has also attracted some attention in the literature.
It was observed in \cite{hevowa96} that 
the relativization result from \cite{bocrsi92,ver93a} {\em does not hold\/}
in the case of unbalanced trees.
Thus, part of the motivation to consider this construct is gone. Nevertheless
definability questions are also interesting in this case. The just
mentioned observation even makes a systematic
comparison of both models a worthwhile study.

\subsubsection{Balanced vs.~Unbalanced Trees}

In \cite{hevowa96} the question of definability of the polynomial hierarchy
was addressed. As mentioned earlier in Theorem~\ref{pollogsigma}, the
classes of the log-time hierarchy exactly define the classes of the
polynomial hierarchy. However, in the case of unbalanced trees, one
can somehow use the tree structure to hide an oracle that is able
to count paths. More formally,

\begin{thm}[\cite{hevowa96}]
$\Leafp{\Siglt{k}} = \left(\Sigp{k}\right)^\PP$.
\end{thm}

\subsubsection{The Acceptance Power of Different Tree Shapes}

Hertrampf \cite{her95} considered besides the above two models also
the definition of classes via leaf languages for computation trees which
are {\em full binary trees.} The obtained classes are noted by
$\FBTLeafp{\cdot}$. 
Though trivially for every $B\seq\Bs$ we have
$\FBTLeafp{B}\seq\BLeafp{B}\seq\Leafp{B}$,
Hertrampf proved the somewhat counterintuitive result, that the definability
power by arbitrary single regular languages  
does not decrease but possibly {\em increases\/} as the tree shapes get 
more and more regular; that is for every regular language $B$ there is a 
regular language $B'$ such that 
$\Leafp{B}=\Leafp{B'}=\BLeafp{B'}$, 
and for every regular language $B$ there is a regular language $B'$ such that 
$\BLeafp{B}=\BLeafp{B'}=\FBTLeafp{B'}$.

\subsubsection{Definability Gaps}
\label{gaps}

In the case of arbitrary tree shapes, Borchert et al.\ were able to
prove the existence of definability gaps. In particular, the following was
shown.

\begin{thm}
Suppose the polynomial hierarchy does not collapse, and let
$B$ be an arbitrary regular language. 
\begin{enumerate} 
\item If $\P\seq\Leafp{B}\seq\K$, then $\Leafp{B}=\P$ or $\Leafp{B}=\K$,
where $\K$ is one of the classes $\NP$, $\co\NP$, or $\Mod_p\P$ 
(for some prime number $p$) {\rm \cite{bor94b}}.
\item If $\NP\seq\Leafp{B}\seq\co\US$, then $\Leafp{B}=\NP$ or 
$\Leafp{B}=\co\US$ {\rm \cite{bokust96}} (analogously for $\co\NP$ and $\US$).
\end{enumerate}
\end{thm}

We come back to questions of this kind in Sect.~\ref{concl}.

\section{Other Resource Bounds}

\subsection{Circuit Classes}
\label{circuitleaf}

Corollary~\ref{pspaceserial} easily yields the following:

\begin{coro}
$\opp{S_5}{\ACn} = \PSPACE$.
\end{coro}

This coincidence between $\opp{\cdot}{\P}$ and $\opp{\cdot}{\ACn}$
holds under more general circumstances. Let $\Neut$ denote the
set of all languages $L\seq\Ss$ that contain a neutral letter $e$, i.e.
for all $u,v\in\Ss$, we have $uv\in L \iff uev\in L$.

\begin{thm}\label{ACvsP}
If $B\in\Neut$ then $\opp{B}{\P} = \opp{B}{\ACn}$.
\end{thm}

\begin{outline}
Correctness of 
computation paths of nondeterministic Turing machines can be checked in $\ACn$
as already pointed out in the proof of Corollary~\ref{pspaceserial}. The
required $\ACn$ computation in input $(x,y)$
now checks that its second input argument
is a correct path of the corresponding machine on input $x$; if so
it outputs $1$ iff this path is accepting and $0$ otherwise. If $y$ does
not encode a correct path then the neutral letter is output.
\end{outline}

A careful inspection of the just given proof reveals that the result not
only holds for language $B\in\Neut$, $B\seq\Bs$, but also for languages
$B$ that are obtained from some $B'\in\Neut$, $B\seq\Ss$ (possibly
$|\Sigma|>2$) by block encoding. The same generalization holds for all
results that we state below for ``$B\in\Neut$'' (i.e., Theorem~\ref{expconp}
and all results in Sect.~\ref{lindst}).

In the context of $\NCe$ and subclasses, some interesting results
can be obtained for classes of the form $\opl{\cdot}{\ACn}$.

First, Barrington's theorem \cite{bar89} yields:

\begin{thm}\label{barrington}
$\opl{S_5}{\ACn} = \NCe$.
\end{thm}

\begin{thm}
\begin{enumerate}
\item $\opl{B}{\ACn} = \NCe$ for every regular language $B$ whose syntactic
mo\-no\-id is non-solvable.
\item $\opl{{\rm SOLVABLE}}{\ACn} = \ACCn$.
\end{enumerate}
\end{thm}

Generally the class $\opl{B}{\ACn}$ roughly corresponds to
$\ACn$ circuits with a $B$ gate on top, e.g.~$\opl{{\rm maj}}{\ACn}$
is the class of all languages accepted by perceptrons.

$\ACn$ circuits with arbitrary $B$ gates are examined in 
\cite{baimst90,baim94} (see also Sect.~\ref{concl}).

\subsection{Logspace and Logtime Leaf Languages}
\label{loglog}

In the same spirit as above for nondeterministic {\em polynomial time\/}
machines, Jenner, McKenzie, and Th\'erien examined in \cite{jemcth94} leaf
languages for nondeterministic {\em logarithmic time\/} and
{\em logarithmic space\/} machines.

First turning to the logspace case, we observe that the trivial way to
formulate $\BLeafl{B}$, the class defined by
logspace machines with leaf language $B$, as a class $\opp{\cdot}{\L}$ does
not work ($\L$ denotes the class of logspace decidable sets). This is because
(for $B\in\Neut$) already $\opp{B}{\P} = \opp{B}{\ACn}$ 
(see Sect.~\ref{circuitleaf}),
and therefore also $\opp{B}{\P} = \opp{B}{\L}$.

However, if we turn to logarithmic space-bounded one-way protocol machines
or 2-1-machines \cite{lan86}, we can come up with a connection.
A {\em 2-1-Turing machine\/} is a Turing machine with two input tapes:
first a (regular) input tape that can be read as often as necessary, 
and second, an additional (protocol) tape that can be read only once 
(from left to right). 
Define \toL\ to be the class of all two argument languages $L$ that can
be computed by logspace-bounded 2-1-TMs such that in the initial configuration,
the first argument of the input is on the regular input tape, and the second
argument is on the one-way input tape.
Then the following can be shown using ideas from \cite{lan86}:

\begin{thm}
Let $B\seq\Bs$. Then $\opp{B}{\toL} = \BLeafl{B}$.
\end{thm}

Jenner, McKenzie, and Th\'erien showed that in a lot of cases,
the balanced and unbalanced model coincide for logarithmic space machines,
and moreover it sometimes coincides with the polynomial time case,
e.g.~Theorem~\ref{pollogsigma} above also holds with  leaf languages
for logspace machines.
Interesting to mention is that in the logarithmic space model,
regular leaf languages define the class $\P$, while $\NCe$ defines the class
$\PSPACE$.

In the logarithmic time case, coincidence with the logarithmic time 
reducibility closure could be shown for all well-behaved leaf languages. 
Formulated in terms of our quantifier, some of their results read as
follows:

\begin{thm}[\cite{jemcth94}]\label{logtimeleaf}
\begin{enumerate}
\item $\opl{\REG}{\DLOGTIME} = \NCe$.
\item $\opl{\CFL}{\DLOGTIME} = \LOGCFL$.
\item $\opl{\CSL}{\DLOGTIME} = \PSPACE$.
\end{enumerate}
\end{thm}

\begin{thm}[\cite{jemcth94}]
\begin{enumerate}
\item $\opl{B}{\DLOGTIME} = \NCe$ for any regular language $B$ whose syntactic
monoid is non-solvable.
\item $\opl{{\rm SOLVABLE}}{\DLOGTIME} = \ACCn$.
\item $\opl{{\rm APERIODIC}}{\DLOGTIME} = \ACn$.
\end{enumerate}
\end{thm}

\subsection{Other models}

\subsubsection{Type 2 Operators}

Operators ranging not over words but over oracles, so called {\em type 2
operators}, have been examined in \cite{bawa96,bovowa96,vowa97b} and elsewhere.
Most of the considered classes coincide with classes of the form
$\op{B}{\F}{\K}$ where $\K=\co\NP$ or $\K=\PSPACE$ and $\F$ is the class of
all exponential time computable functions (let us write
$\ope{B}{\K}$ as a shorthand for this choice of $\F$).
A word of care about the computational model however is in order now.
We say that a language $L$ belongs to the class $\ope{B}{\co\NP}$ if 
there is a function $f$ computable in exponential time, and a 
set $A$ such that $x\in L \iff \x{A_x}{0}{f(x)}\in B$, where $A$ is 
accepted by some co-nondeterministic Turing machine $M$ 
that on input $\langle x,y\rangle$ runs in time polynomial in the length
of $x$. The length of $y$ is possibly exponential in the length of $x$;
thus to enable $M$ to access all positions of $y$ within its time bound
we supply $M$ with a regular input tape on which $x$ is found, and
a second input tape for $y$, which is accessed by an index tape . This special
input tape is similar to an oracle tape, and therefore quantifiers over
strings on this tape translate to quantifiers over oracles.
(In the case of $\ope{B}{\PSPACE}$ we require our machines to use space
no more than polynomial in the length of their regular input $x$.)

\begin{thm}\label{expconp}
Let $B\in\Neut$. Then we have: 
$$\ope{B}{\EXPTIME} = \ope{B}{\PSPACE} = \ope{B}{\co\NP}.$$
\end{thm}

\begin{outline}
If we look at the proof of Theorem~\ref{ACvsP} we see that to check correct
computation paths we actually don't need the full power of $\ACn$. 
$\Pilt{1}$ is sufficient, but we have to modify the computation model
slightly as follows: The log-time machine has a regular input tape (which
is accessed as usual by using an index tape) and a second input tape on which
the path to be checked is given (again access is by an index tape).
We thus get:
$$\opp{B}{\P}=\opp{B}{\ACn}=\opp{B}{\Pilt{1}}.$$
Using standard translation arguments we now get the claim of the theorem
by lifting up this equation one exponential level.
\end{outline}

\subsubsection{\protect{$\NCe$} Leaf Languages}

In \cite{camcthvo96} leaf languages for nondeterministic finite automata
were considered. The original input is however first given into a uniform
projection, and the result of this projection is then fed into the NFA.
Barrington's Theorem \ref{barrington} implies that with regular leaf
languages we thus get exactly the class $\NCe$. Some other characterizations
were given in \cite{camcthvo96}, and the model was also used to examine
counting classes within $\NCe$.

\subsubsection{Function Classes}

In \cite{koscvo97} the definability of function classes has been
examined. An oracle separation criterion generalizing Theorem~\ref{septhm}
was given and applied successfully in some open cases.

\section{Leaf Languages vs.~Lindstr\"om Quantifiers}
\label{lindst}

Lindstr\"om quantifiers \cite{lin66} are a well established generalized
quantifier notion in finite model theory. The reader probably has noticed
some resemblance of our definition in Sect.~\ref{defin} with that
of Lindstr\"om quantifiers. It will be our aim in the upcoming sections
to make this precise.

As we will see there is a strong connection between leaf languages
for polynomial time machines and {\em second-order\/} Lindstr\"om
quantifiers. Since this notion might not be so well-known, we give---after
very briefly recalling some terminology from finite model theory---a
precise definition in Sect.~\ref{secondorder}.

In later subsections we will have the need to talk about {\em the
second-order version of a given first-order Lindstr\"om quantifier.}
We chose to make this precise by talking about the semantics of
quantifiers given by {\em languages\/} instead of the usual way of
defining semantics by classes of structures. In the next subsection,
we will define how a language $B$ gives rise to a first-order
quantifier $\Qn$ and a second order quantifier $\Qe$.

\subsection{Second-Order Lindstr\"om Quantifiers}
\label{secondorder}

A signature is a finite sequence 
$\tau = \langle R_1,\dots,R_k,c_1,\dots,c_\ell\rangle$
of relation symbols and constant symbols.
A finite structure of signature $\tau$ is a tuple
$\calA = (A, R_1^\calA,\dots,R_k^\calA,\allowbreak c_1^\calA,
\dots,c_\ell^\calA)$
consisting of a finite set $A$ (the universe of $\calA$) and
interpretations of the symbols in $\tau$ by relations over $\calA$ (of
appropriate arity) and elements of $\calA$. 
$\Struct{\tau}$ is the set of all {\em finite ordered structures\/} over $\tau$.
The {\em characteristic string $\chi_R$} of a relation 
$R\in\{0,\dots,n-1\}^a$
is the string $\chi_R\eqdef b_1\cdots b_{n^a}$ where $b_i=1$ iff 
the $i$-th vector in $\{0,\dots,n-1\}^a$ 
(in the order $(0,\dots,0,0)<(0,\dots,0,1)<(n-1,\dots,n-1,n-1)$)
is in $R$. For $1\leq i\leq n^a$, let $\chi_R[i]$ denote the $i$-th bit
in $\chi_R$.

%
If $\calL$ is a logic (as e.g.~$\FO$ or $\SO$) and $\K$ is a complexity
class, then we say that {\em $\calL$ captures $\K$\/} if every property
over (standard encodings of) structures decidable within $\K$ is expressible by
$\calL$ sentences, and on the other hand for every fixed $\calL$ sentence
$\phi$, determining whether $\calA\models\phi$ can be done in $\K$.
As an abbreviation we will most of the time simply write $\K=\calL$.

A first-order formula $\phi$ with $k$ free variables defines for every
structure $\calA$ the relation 
$\phi^\calA\eqdef\set{\vec{a}\in A^k}{\calA\models\phi(\vec{a})}$,
see \cite{ebfl95}.

Every class of structures $K\seq\Struct{\sigma}$ 
over a signature $\sigma=\langle P_1,\dots,\P_s\rangle$ defines the
first-order Lindstr\"om quantifier $Q_K$ as follows:
Let $\phi_1,\dots,\phi_s$ be first-order formulae over signature $\tau$
such that for $1\leq i\leq s$ the number of
free variables in $\phi_i$ is equal to the arity of $P_i$. Then
$$Q_K\vec{x}_1,\dots,\vec{x}_s\left[\phi_1(\vec{x}_1),
\dots,\phi_s(\vec{x}_s)\right]$$ 
is a 
$Q_K\FO$ formula. If $\calA\in\Struct{\tau}$, then
$$\calA \models Q_K\vec{x}_1,\dots,\vec{x}_s\left[\phi_1(\vec{x}_1),
\dots,\phi_s(\vec{x}_s)\right]$$
iff $(A,\phi_1^\calA,\dots,\phi_s^\calA)\in K$.

The just given definition is the original definition given by
Lindstr\"om \cite{lin66}, which the reader will also find in
textbooks, see e.g.~\cite{ebb85,ebfl95}.
For our examinations, the following equivalent
formulation will be useful (observe that this only makes sense for
ordered structures):

Given a first-order formula $\phi$ with $k$ free variables and a corresponding
finite ordered structure $\calA$, this defines the binary string 
$\chi_{\phi^\calA}$ of length $n^k$ ($n=|A|$).
Now given a sequence $\phi_1,\dots,\phi_s$ of formulae with $k$ free
variables each and a structure $\calA$, we similarly get the
tuple $(\chi_{\phi_1^\calA},\dots,\chi_{\phi_s^\calA})$,
where $|\chi_{\phi_1^\calA}|=\cdots=|\chi_{\phi_s^\calA}| = n^k$.
Certainly, there is a one-one correspondence between such tuples and 
strings of length $n^k$ over a larger alphabet (in our case with $2^s$ 
elements) as follows. Let $A_s$ be such an alphabet. Fix an arbitrary
enumeration of $A_s$, i.e.~$A_s=\{a_0,a_1,\dots,a_{2^s-1}\}$.
Then  $(\chi_{\phi_1^\calA},\dots,\chi_{\phi_s^\calA})$ corresponds to
the string $b_1b_2\cdots b_{n^k}$, where for $1\leq i\leq n^k$,
$b_i\in A_s$,
$b_i=a_k$ for that $k$ whose length $s$ binary representation
(possibly with leading zeroes) is given by
$\chi_{\phi_1^\calA}[i]\cdots\chi_{\phi_s^\calA}[i]$.
In symbols: 
$w_s(\chi_{\phi_1^\calA},\dots,\chi_{\phi_s^\calA}) = b_1b_2\cdots b_{n^k}$.

This leads us to the following definition: 
A sequence $\left[\phi_1,\dots,\phi_s\right]$ is in {\em first-order
word normal form,} 
iff the $\phi_i$ have the same number $k$ of
free variables. Let $\Gamma$ be an alphabet such that $|\Gamma|\geq2^s$,
and let $B\seq\Gamma^*$. Then 
$\calA \models Q_B\vec{x}\left[\phi_1(\vec{x}),\dots,\phi_s(\vec{x})\right]$
iff 
$w_s(\chi_{\phi_1^\calA},\dots,\chi_{\phi_s^\calA}) \in B$.

It can be shown \cite{buvo98,bur96} that every Lindstr\"om quantifier $Q_K$ 
can without loss of generality be assumed to be of the form $Q_B$ as just 
defined. This is the case since for every sequence 
$\left[\phi_1,\dots,\phi_s\right]$ of first-order formulae we find
an equivalent sequence in word normal form such that the corresponding
formulae with Lindstr\"om quantifier express the same property.

{\em Second-order Lindstr\"om quantifiers\/} are defined as follows
\cite{buvo98,bur96}:
Given a formula $\phi$ with free second-order variables $P_1,\dots,P_m$
and a structure $\calA$, define
$\phi^{2^\calA}\eqdef\set{(R_1^\calA,\dots,R_m^\calA)}{\calA\models
\phi(R_1^\calA,\dots,R_m^\calA)}$, and let
$\chi_{\phi^{2^\calA}}$ be the corresponding characteristic string,
the order of vectors of relations being the natural one induced by
the underlying order of the universe. If the arities of
$P_1,\dots,P_m$ are $r_1,\dots,r_m$, resp., then the length of 
$\chi_{\phi^{2^\calA}}$ is 
$2^{n^{r_1}+\cdots+n^{r_m}}$ 

Let $\sigma=\langle \sigma_1,\dots,\sigma_s\rangle$ be a signature,
where $\sigma_i=\langle P_{i,1},\dots,P_{i,m_i}\rangle$
for $1\leq i\leq s$. Thus $\sigma$
is a signature consisting of a sequence of $s$ signatures with only predicate
symbols each. Let $\ell_{i,j}$ be the arity of $P_{i,j}$. 
A {\em second-order structure\/} of signature $\sigma$
is a tuple $\calA=(A,\CR_1,\dots,\CR_s)$, where for every $1\leq i\leq s$,
$\CR_i\seq\set{(R_{i,1},\dots,R_{i,m_i})}{R_{i,j}\seq A^{\ell_{i,j}}}$.
Given now a signature $\tau$ and second-order formulae 
$\phi_1(\vec{X}_1),\dots,\phi_s(\vec{X}_s)$ over $\tau$ where for every
$1\leq i\leq s$ the number and arity of free predicates in $\phi_i$
corresponds to $\sigma_i$. Let $\K$ be a class of second-order structures
over $\sigma$. Then
$Q_\K \vec{X}_1,\dots,\vec{X}_s\left[\phi_1(\vec{X}_1),
\dots,\phi_s(\vec{X}_s)\right]$
is a $Q_\K\SO$ formula. If $\calA\in\Struct{\tau}$, then
$\calA\models Q_\K \vec{X}_1,\dots,\vec{X}_s\left[\phi_1(\vec{X}_1),
\dots,\phi_s(\vec{X}_s)\right]$ iff
$(A,\phi_1^{2^\calA},\dots,\phi_s^{2^\calA})\in\K$.

Again, we want to talk about second-order Lindstr\"om quantifiers 
defined by languages. Thus we define analogously to the above:
A sequence $\left[\phi_1(\vec{X}_1),\dots,\phi_s(\vec{X}_s)\right]$ of
second-order formulae is in {\em second-order word normal form}, if
the $\phi_1,\dots,\phi_s$ have the same predicate symbols,
i.e.~in the above terminology 
$\sigma_1=\cdots=\sigma_s=\langle P_1,\dots,P_m\rangle$.
Let for $1\leq i\leq m$ the arity of $P_i$ be $r_i$.
Observe that in this case, 
$|\chi_{\phi_1^{2^\calA}}|=\cdots=|\chi_{\phi_s^{2^\calA}}|
= 2^{n^{r_1}+\cdots+n^{r_m}}$ (for $n=|A|$),
thus $(\chi_{\phi_1^{2^\calA}},\dots,\chi_{\phi_s^{2^\calA}})$ 
corresponds to a word of the same length over an alphabet of cardinality $2^s$.
Given now a language $B\seq\Gamma^*$ with $|\Gamma|\geq2^s$,
the second-order Lindstr\"om quantifier given by $B$ is defined by
$\calA\models 
Q^1_B\vec{X}\left[\phi_1(\vec{X}),\dots,\phi_s(\vec{X})\right]$
iff 
$w_s(\chi_{\phi_1^{2^\calA}},\dots,\chi_{\phi_s^{2^\calA}}) \in B$.

Again it was shown in \cite{buvo98,bur96} that for every second-order
Lindstr\"om quantifier $Q_\K$ there is an equivalent $Q^1_B$.

When talking about the first-order Lindstr\"om quantifier given by $B$,
we sometimes explicitly write $Q^0_B$ instead of $Q_B$.
In addition to the above logics $\Qn\FO$ and $\Qe\SO$ where we allow
Lindstr\"om quantifiers followed by an arbitrary first-order (second-order,
resp.) formula, we also need $\Qe\FO$ (where we have a second-order
Lindstr\"om quantifier followed
by a formula with no other second-order quantifiers), and
$\FO(\Qn)$ and $\SO(\Qe)$ (where we have first-order (second-order, resp.)
formulae with arbitrary nesting of universal, existential, and Lindstr\"om
quantifiers). 
For a class of languages $\CC$ we use the notation $Q_\CC$ with the obvious
meaning, e.g.~$\FO(Q^0_\CC)$ denotes all first-order sentences with arbitrary
quantifiers $\Qn$ for $B\in\CC$.

\subsection{A Logical Characterization of the Leaf Concept}

The main technical connection between polynomial time leaf languages
and second-order Lindstr\"om quantifiers is given in the following
theorem:

\begin{thm}
Let $M$ be a polynomial time nondeterministic machine whose
computation tree is always a full binary tree, and 
let $B\seq\Bs$. Then there is a
$\Sigma^1_1$ formula $\phi$ such that
$$\LeafM{B} = \Qe\vec{X}\left[\phi(\vec{X})\right].$$
\end{thm}

\begin{outline}
We use a modification of Fagin's proof \cite{fag74}. The 
$\Qe$ quantifier will bind the nondeterministic guesses of the machine.
The second-order quantifiers in $\phi$ will bind variables $Y$ that
encode computation paths of $M$. The formula $\phi(X)$ says
``there is a $Y$ encoding a correct computation path of $M$ corresponding
to nondeterministic guesses $X$, which is accepting.''
\end{outline}

If we deal with $B\seq\Gs$ not necessarily over the binary alphabet, then
instead of $\phi$ above, we get formulae $\phi_s,\dots,\phi_s$ such that
$$\LeafM{B} = \Qe\vec{X}\left[\phi_1(\vec{X}),\dots,\phi_s(\vec{X})\right].$$
$\phi_i(X)$ says ``there is a $Y$ encoding a correct computation path of $M$ 
corresponding to nondeterministic guesses $X$, and the leaf symbol produced
on this path has a $1$ in bit position $i$ (in binary). Thus what we have
here is some block-encoding of $\Gamma$ in binary strings of length $s$.

The just given theorem shows that $\FBTLeafp{B}\seq \Qe\Sigma^1_1$.
The question now of course is if there is a logic capturing $\FBTLeafp{B}$.
For the special case $B\in\Neut$, the answer is yes.

\begin{thm}[\cite{buvo98}]\label{modelleaf}
Let $B\in\Neut$. Then $\Qe\FO = \BLeafp{B}$.
\end{thm}

\begin{outline}
This time $\Qe$ binds the nondeterministic guesses $X$ as well as the
encoding $Y$ of a possible computation path. The first order formulae
``output'' the neutral letter, if $Y$ does not encode a correct path.
This proves the direction from right to left. For the other inclusion,
we observe that we can design a Turing machine which branches on all possible
assignments for the relational variables and then simply evaluates the
first-order part.
\end{outline}

In the preceding theorem $\BLeafp{B}$ is captured by the logic
$\Qe\FO$ {\em uniformly\/} in the sense of \cite{mapn93,mapn94};
this means that the particular formula describing the Turing machine
is independent of the leaf language.

Let us next address the question if the quantifier in the preceding theorem
is genuinely second-order. First, we have to give some definitions.
A succinct representation \cite{wag86b,baloto92,vei96}
of a binary word $x$ is a boolean circuit giving on input $i$ the $i$th bit 
of $x$. The succinct version $sA$ of a language $A$ is the following: 
Given a boolean circuit describing a word $x$, 
is $x=x_10x_20\cdots x_{n-1}0x_n1w$ for arbitrary $w\in\{0,1\}^*$, such
that $x_1x_2\cdots x_n \in A$?
The boolean circuits we allow are standard unbounded fan-in circuits
over AND, OR, NOT. The encoding consists of a sequence of tuples
$(g,t,h)$, where $g$ and $h$ are gates, $t$ is the type
of $g$, and $h$ is an input gate to $g$ (if $g$ is not already an 
input variable).

Now we see that there is an equivalent first-order logic for $\Qe\FO$.

\begin{thm}
Let $B\in\Neut$. Then $\BLeafp{B} = \Qe\FO = Q^0_{sB}\FO$.
\end{thm}

\begin{outline}
Veith \cite{vei96} showed that $sB$ is complete for $\BLeafp{B}$ under
projection reductions. (A somewhat weaker result appeared in \cite{bolo96}). 
This together with Theorem~\ref{modelleaf} implies the theorem.
\end{outline}

\subsection{Applications}

Burtschick and Vollmer in \cite{buvo98} also examined logically defined
leaf languages. It turned out that if the leaf language is given by a
first-order formula, then the obtained complexity class is captured
by the corresponding second-order logic. More specifically, they
proved for instance:

\begin{thm}[\cite{buvo98}]\label{leaflindfo} 
Let $B\in\Neut$. Then $\oppP{\Qn\Sigma^0_k} = \Qe\Sigma^1_k$.
\end{thm}

As a special case of Theorem~\ref{leaflindfo} we get a characterization
of the classes of the polynomial hierarchy which is tighter than the
one in Theorem~\ref{pollogsigma}.

\begin{coro}
$\oppP{\Sigma^0_k} = \Sigp{k}$.
\end{coro}

From the $\PSPACE$ characterization Theorem~\ref{pspace} and the above
results, we get the following model-theoretic characterization of $\PSPACE$:

\begin{coro}
$Q^1_{S_5}\FO = Q^0_{sS_5}\FO = \PSPACE$.
\end{coro}

\subsection{First-order quantifiers}
\label{logicother}

It is known from the work of Immerman et al.\ \cite{imm89,baimst90} that
(uniform) $\ACn$ is captured by $\FO$. However, for this result, we have
to include the bit predicate in our logic. We make this assumption throughout
this subsection (all the previously given 
results are valid without the bit predicate).

\begin{thm}\label{bit}
Let $B\seq\Bs$. Then $\opl{B}{\ACn} = \Qn\FO$.
\end{thm}

Theorem~\ref{bit}, together with results from Sect.~\ref{loglog} on
logtime leaf languages, gives some more model-theoretic characterizations.

\begin{coro}\label{bitcharakt}
\begin{enumerate}
\item $\PSPACE = Q_{\CSL}^0\FO = \FO(Q_{\CSL})$.
\item $\LOGCFL = Q_{\CFL}^0\FO = \FO(Q_{\CFL})$.
\end{enumerate}
\end{coro}

\begin{outline}
One can show that generally $\BLeaflt{B}\seq\Qn\FO$. The corollary then
follows from Theorem~\ref{logtimeleaf}.
\end{outline}

\section{Conclusion}
\label{concl}

We examined a generalized quantifier notion in computational complexity.
We proved that not only all quantifiers examined so far (whether in the
logarithmic, polynomial, or exponential time context) can be seen as
special cases of this quantifier, but also circuits with generalized gates
and Turing machines with leaf language acceptance.

Most of the emerging complexity classes can be characterized by means from
finite model theory. We gave a precise connection to finite model theory
by showing how complexity classes defined by the generalized quantifier
relate to classes of finite models defined by logics enhanced with Lindstr\"om 
quantifiers. 

A number of questions remain open. The results we gave in Sect.~\ref{lindst}
related complexity classes to logics of the form ``Lindstr\"om quantifier
followed by a usual first- or second-order formula.'' It is not clear
if logics defined by arbitrary nesting of Lindstr\"om quantifiers have
a nice equivalent in terms of the generalized complexity theoretic quantifier.
Barrington, Immerman, and Straubing proved:

\begin{thm}[\cite{baimst90}]
Let $B\in\Neut$. Then $\FO(\Qn) = \ACn[B]$ ($\ACn$ circuits with $B$ gates).
\end{thm}

Moreover one can show:

\begin{thm}[\cite{vol96b}]
Let $B\in\Neut$. Then
$\FO(\Qe)$ is the oracle hierarchy given by $\opp{B}{\ACn}$ as building block.
\end{thm}

But the general relationship remains unclear. 
The work of Makowsky and Pnueli (see \cite{mapn93,mapn94}), Stewart
(see e.g.~\cite{ste91,ste92}), and Gottlob (see\cite{got95}) shows that
there is a strong relation between Lindstr\"om logics and relativized
computation. The just mentioned results also hint in that direction.
Gottlob \cite{got95} related the expressive power of logics of the
form ``Lindstr\"om quantifier Q followed by first-order formula'' to
the expressive power of $\FO(Q)$. However his results only apply for
superclasses of $\L$ (logarithmic space). Interesting cases within $\NCe$
remain open. Generally the connection between prenex Lindstr\"om logics vs.\ 
logics allowing arbitrary quantifier nestings on the model theoretic side, and
leaf languages vs.\ oracle computations on the complexity theoretic side
should be made clearer.

It is open for which of the results in Sect.~\ref{logicother} the bit 
predicate is really needed. One can show that without bit,
$Q_{\CFL}\FO = \CFL$ contrasting the corresponding result with bit given
in Corollary~\ref{bitcharakt}.
The power of the bit predicate in this context deserves further attention.

From a complexity theoretic point of view, we think the main open question
is the following. A lot of classes defined by leaf languages have been
identified. However, most of the results are not about singular leaf
languages but about classes of leaf languages. For example (see
Theorem~\ref{algebraic}), 
if we take an arbitrary aperiodic leaf languages, then 
the complexity class we obtain is included in $\PH$, and conversely
we get all of $\PH$ when we allow aperiodic leaf languages:
$\BLeafp{{\rm APERIODIC}}=\PH$. The question
now is the following: What exactly are the classes of the form 
$\BLeafp{B}$ for aperiodic $B$? Is it possible to come up with a complete
list of classes that can be defined in this way? 
Some of the results in Sect.~\ref{gaps} point in this direction. For
example we know that there is no class between $\P$ and $\NP$ that can
be defined by a regular leaf language (unfortunately the result given
in Sect.~\ref{gaps} holds only for the unbalanced case). Can we come
up with similar result for the balanced case? Generally, very little
is known about the power of {\em single\/} leaf languages as opposed to
classes of leaf languages.

\bigskip\par\noindent{\bf Acknowledgment.}
For helpful discussions I am grateful to J.~Makowsky (Haifa) and
H.~Schmitz (W\"urzburg).

\bibliography{komp}
\bibliographystyle{alpha}

\end{document}